\pdfoutput=1

\documentclass[11pt,twoside,a4paper,cmspaper,final,collab]{cms-tdr}
\def\svnVersion{266112}\def\cmsCernNo{2013-037}\def\cmsCernDate{\today}\def\cmsMessage{Published in Nature Physics as \href{http://dx.doi.org/10.1038/nphys3005}{\doi{10.1038/nphys3005.}}}

\begin{document}\cmsNoteHeader{HIG-13-033}

\hyphenation{had-ron-i-za-tion}
\hyphenation{cal-or-i-me-ter}
\hyphenation{de-vices}

\RCS$Revision: 266112 $
\RCS$HeadURL: svn+ssh://alverson@svn.cern.ch/reps/tdr2/papers/HIG-13-033/trunk/HIG-13-033.tex $
\RCS$Id: HIG-13-033.tex 266112 2014-11-02 14:41:24Z alverson $
\ifthenelse{\boolean{cms@external}}{\newsavebox{\CMSabstractboxN}\renewcommand{\abstractname}{\relax}\renewcommand{\abstract}[1]{\savebox{\CMSabstractboxN}{\parbox[t]{\textwidth}{\textbf{#1}}}}\savebox{\CMSabstractbox}{}}{} 
\newcommand{\hbb}{\ensuremath{\PH\rightarrow\bbbar}\xspace}
\newcommand{\htt}{\ensuremath{\PH\rightarrow\Pgt\Pgt}\xspace}
\newcommand{\tautau}{\ensuremath{\Pgt\Pgt}\xspace}
\newcommand{\hww}{\ensuremath{\PH\rightarrow\PW\PW}\xspace}
\newcommand{\ww}{\ensuremath{\PW\PW}\xspace}
\newcommand{\emu}{\ensuremath{\Pe\Pgm}\xspace}
\newcommand{\mh}{\ensuremath{m_{\PH}}\xspace}
\newcommand{\mhp}{\ensuremath{m_{\PH}^{\prime}}\xspace}
\newlength{\cmsFigWidth}
\makeatletter\if@twocolumn
\setlength{\cmsFigWidth}{0.88\columnwidth}
\else
\setlength{\cmsFigWidth}{0.70\columnwidth}
\fi\makeatother
\cmsNoteHeader{HIG-13-033} 

\title{Evidence for the direct decay of the 125\GeV Higgs boson to fermions}

\date{\today}

\abstract{
The discovery of a new boson with a mass of approximately 125\GeV in 2012 at
the LHC~\cite{Chatrchyan:2012ufa,Chatrchyan:2013lba,Aad:2012tfa}
 has heralded a new era in understanding the nature of electroweak
symmetry breaking
and possibly completing the standard
 model of particle physics~\cite{Englert:1964et,Higgs:1964ia,Higgs:1964pj,Guralnik:1964eu,Higgs:1966ev,
Kibble:1967sv}.
Since the first observation in decays to $\cPgg\cPgg$, $\PW\PW$, and
$\cPZ\cPZ$ boson pairs, an extensive set of measurements of the
mass~\cite{Chatrchyan:2012jja,Chatrchyan:2013mxa} and couplings to
$\PW$ and $\cPZ$
bosons~\cite{Aad:2013wqa,Chatrchyan:2013iaa,Chatrchyan:2013mxa}, as well as
multiple tests of the spin-parity quantum
numbers~\cite{Chatrchyan:2012jja,Aad:2013xqa,Chatrchyan:2013iaa,Chatrchyan:2013mxa},
have revealed that the properties of the new boson are consistent with those of the
long-sought agent responsible for electroweak symmetry breaking.
An important open question is whether the new
particle also couples to fermions, and in particular to down-type fermions,
since the current measurements
mainly constrain the couplings to the up-type top quark.
Determination of the couplings to down-type fermions requires direct
measurement of the corresponding Higgs boson decays, as recently
reported by the CMS experiment in the study of Higgs decays to bottom
quarks~\cite{Chatrchyan:2013zna} and $\Pgt$ leptons~\cite{Chatrchyan:2014nva}.
In this letter we report the combination of these two channels which results, for
the first time, in strong evidence for the direct coupling of the 125\GeV Higgs
boson to down-type fermions, with an observed significance of
3.8 standard deviations, when 4.4 are expected.
}

\hypersetup{%
pdfauthor={CMS Collaboration},%
pdftitle={Evidence for the direct decay of the 125 GeV Higgs boson to
fermions},%
pdfsubject={CMS},%
pdfkeywords={CMS, physics, Higgs, Yukawa coupling}}

\maketitle 

\ifthenelse{\boolean{cms@external}}{\textbf{\unhbox\CMSabstractboxN}}{}

The CMS and ATLAS  experiments at the Large Hadron
Collider (LHC) have reported the discovery of a new
boson~\cite{Chatrchyan:2012ufa,Chatrchyan:2013lba,Aad:2012tfa} with a
mass near
125\GeV and with production rates, decay rates, and spin-parity quantum
numbers~\cite{Chatrchyan:2012jja,Aad:2013wqa,Aad:2013xqa,Chatrchyan:2013iaa,Chatrchyan:2013mxa}
compatible with those expected for the standard model Higgs
boson~\cite{Englert:1964et,Higgs:1964ia,Higgs:1964pj,Guralnik:1964eu,Higgs:1966ev,
Kibble:1967sv}.
In the standard model, the Higgs boson is a spin zero particle predicted to arise from the
Higgs field which is responsible for electroweak symmetry
breaking~\cite{Glashow1961579,PhysRevLett.19.1264}.
As such, the standard model Higgs boson couples directly to the $\PW$ and
$\cPZ$ bosons, and indirectly to photons.
To date, significant signals have been reported in channels where the boson
decays to either $\cPgg\cPgg$, $\ww$, or $\cPZ\cPZ$ boson
pairs~\cite{Aad:2013wqa,Chatrchyan:2013iaa,Chatrchyan:2013mxa} as predicted by
the theory.
Overall, these results directly demonstrate that the new particle is intimately
related to the mechanism of spontaneous electroweak symmetry breaking, whereby the $\PW$ and
$\cPZ$ bosons become massive, and thus it has been identified as a Higgs boson.

The standard model also predicts that the Higgs field couples to
fermions through a Yukawa interaction,
giving rise to the masses of quarks and leptons.
The structure of the Yukawa interaction is such that the coupling strength
between the standard model Higgs boson and a fermion is proportional to the
mass of that fermion.
As the masses of many quarks and leptons are sufficiently well known
from experiment, it is possible within the standard model to accurately predict the Higgs
boson decay rates to these fermions.
The existence of such decays and the corresponding rates
remain to be established and measured by experiment.
Indirect evidence for the Higgs coupling to the top quark, an up-type quark
and heaviest elementary particle known to date, is implied by an
overall agreement of the gluon-gluon fusion production channel cross section with the standard model prediction.
However, the masses of down-type fermions may come about through
different mechanisms in theories beyond the standard
model~\cite{RPP2013Higgs}.
Therefore, it is imperative to observe the direct decay of this new
particle to down-type fermions to
firmly establish its nature.
As a consequence of the Yukawa interaction discussed above, the most abundant
fermionic Higgs boson decays will be to third-generation quarks and leptons,
namely the bottom quark and the $\Pgt$ lepton, since the decay of a Higgs boson
with a mass around 125\GeV to top quarks is kinematically not allowed.
Therefore, the most promising experimental avenue to explore
the direct coupling of the standard model Higgs boson to fermions is in the
study of the decay to bottom quark-antiquark pairs
(denoted as $\bbbar$) as well as to tau
lepton-antilepton pairs (denoted as $\tautau$).

Recently, the CMS Collaboration reported on a search for the decays of the
new boson to $\bbbar$ quark pairs~\cite{Chatrchyan:2013zna} as well as to
$\tautau$ lepton pairs~\cite{Chatrchyan:2014nva} based on
data collected in 2011 and 2012.
In this letter, we report on the combination of the results from the study
of these two decays to down-type fermion-antifermion pairs, performed for the
first time at the LHC.

The CMS apparatus comprises several detectors specialised in
identifying different types of particles.
These detectors are arranged inside and outside a superconducting
solenoid of 6\unit{m} internal diameter, that provides a magnetic field of
3.8\unit{T}.
The detector electronics process collision information at a rate of
up to 40\unit{MHz}, and decide whether or not the crossing of proton beam
bunches that took place at the centre of the detector produced proton-proton
collisions of sufficient interest.
Through a layered decision system, from the 20\unit{MHz} of proton
bunch crossings, fewer than 1\unit{kHz} are saved for further analysis.
A detailed description of the full CMS apparatus can be found in
Ref.~\cite{Chatrchyan:2008aa}.
Data were collected at the LHC which delivered proton-proton collisions at
centre-of-mass energies of 7\TeV (2011) and 8\TeV (2012).
A total integrated luminosity of 5.1 (4.9)\fbinv and 18.9 (19.7)\fbinv,
for the \hbb ($\htt$) decay, has been analyzed at 7 and 8 TeV,
respectively.
Each proton bunch crossing gave rise to a large number of
simultaneous proton-proton collisions, on average 9 in 2011 and 21 in 2012.
Such a large number of overlapping collisions presented exceptional
challenges in
reconstructing the individual particles produced in the collision where an
interesting interaction took place.
Those challenges were successfully met thanks to the CMS tracking
system, able to separate collision vertices as close as 0.5\mm
along the beam direction.

The Higgs boson lifetime is ${\sim}10^{-22}$ seconds.
As a consequence, the
detectors at the LHC only record the interactions of its decay products.
For a Higgs boson mass of 125\GeV, the channels expected to be experimentally
accessible include the decays to two photons, two $\PW$ or $\cPZ$ bosons, a
$\bbbar$ quark pair, and a $\tautau$ lepton pair.
The last two are especially challenging since the backgrounds due to
standard model processes other than Higgs boson production are overwhelming and
the invariant mass distributions of the reconstructed decay products do not
produce sharp peaks.
Techniques to estimate the background contributions directly from
the collision data have been developed to face this challenge.
The results are typically expressed in terms of upper limits on the product of
the production and decay rates, significances for the background-only hypothesis to
describe excesses as large, or larger, than those present in the data, and signal strengths relative to the expectation for the
production and decay of the standard model Higgs boson.
Both the observations in data and the expectations for the standard model Higgs boson are
quoted.
The expected quantities are important measures of the sensitivity of a
particular search which do not depend on statistical fluctuations affecting
actual observations in the data.
Thus, they can also be used to compare the sensitivity of different experiments.

The CMS search for a Higgs boson decaying
into a $\bbbar$ quark pair is reported
in Ref.~\cite{Chatrchyan:2013zna}.
The inclusive production of Higgs bosons in this decay channel is extremely hard
to exploit due to the overwhelming direct production of $\bbbar$ quark pairs, as
predicted by quantum chromodynamics (QCD).
For this reason, present searches target the so-called ``VH production'' mode in
which the Higgs boson is produced in association with a heavy vector boson,
V, namely a $\PW$ or a $\cPZ$ boson, identified via its leptonic decay.
Requiring the presence of additional leptons with high transverse momentum
provides an effective handle to reduce the QCD background.
The Higgs boson candidate is reconstructed from two jets of particles, each of
which has been identified as having a high probability of arising from the hadronisation of a bottom quark.
The identification is based on properties of hadrons containing
bottom quarks, such as their long lifetimes which lead to displaced
vertices, or their decay to muons, both producing
distinctive signatures that the CMS detector is sensitive to.
Owing to the 10\% mass resolution of the identified pair of
b-quark jets~\cite{Chatrchyan:2013zna}, the standard model Higgs boson signal is expected to
produce a broad enhancement over the background in the $\bbbar$ invariant mass
distribution.
The result of this analysis shows an excess over the background
expectation for Higgs boson mass hypotheses, \mh, in the range
of 120--135\GeV, with an observed (expected for the standard model Higgs boson) significance
of 2.1 (2.1) standard deviations, $\sigma$, at a Higgs boson mass of
125\GeV.
Correspondingly, the strength of this excess relative to the
expectation for the standard model Higgs boson is $1.0 \pm 0.5$, where the
uncertainty includes the effect of both statistical and systematic
sources of uncertainty.
A result obtained by the CDF and D0 experiments at
the Tevatron proton-antiproton collider, which operated at a centre-of-mass
collision energy of 1.96\TeV, initially revealed an excess of $\bbbar$ events
consistent with the mass of the Higgs boson observed at the LHC, yielding an
observed significance of $2.8\sigma$ for $\mh =
125\GeV$~\cite{Aaltonen:2012qt}.
In a subsequent analysis the observed significance decreased, with the
CDF and D0 experiments reporting an excess $1.59^{+0.69}_{-0.72}$ times the
expectation for the standard model Higgs boson with $\mh =
125\GeV$~\cite{PhysRevD.88.052014}.
At the LHC, in previous studies based solely on the 2011 data, the CMS and ATLAS
Collaborations have reported 95\% confidence level (CL) upper limits of 2.5 to
5.5 times the standard model expectation for a Higgs boson in the 110--130\GeV mass
range~\cite{Chatrchyan:2012ww,Aad:2012gxa}.

Evidence for a Higgs boson decaying to a $\tautau$ lepton pair is
reported by the CMS Collaboration in Ref.~\cite{Chatrchyan:2014nva}.
The lifetime of the $\Pgt$ lepton is $2.9\ten{-13}\unit{s}$, short enough that
$\Pgt$ leptons decay before reaching any of the CMS detector systems.
The predominant decay modes of $\Pgt$ leptons are to a lighter lepton, \ie an
electron or a muon, and two neutrinos, or to one or more pions and a neutrino~\cite{Beringer:1900zz}.
In all cases, among the decay products there are neutrinos which
cannot be directly detected.
However, the presence of neutrinos among the products of an interaction can
be inferred from an overall imbalance in the transverse momentum, measured
using the visible products of the interaction.
The $\Pgt$-lepton decays involving pions are dominated by the decays to a
single charged pion, a charged and a neutral pion, or three charged pions
observed as a collimated jet of particles.
All of these decay modes can be identified by the CMS detector, providing
an important handle to suppress the QCD background.
The main irreducible background is due to the copiously
produced $\cPZ$ bosons which, like the Higgs boson, can decay to a
$\tautau$ lepton pair.
The mass difference between the $\cPZ$ boson (91.2\GeV) and the region where
a Higgs boson was observed (${\sim}125\GeV$), together with a dedicated
$\tautau$ mass reconstruction technique~\cite{Chatrchyan:2014nva},
allow for a good separation of the two contributions.
The gluon-gluon fusion production, the associated V\PH production, and the
vector-boson fusion production processes are probed through the classification of the
events according to kinematic and topological properties of the reconstructed
particles.
The result of the analysis shows a broad excess over background in the Higgs
boson mass range of 115--130\GeV with an observed (expected)
significance of $3.2\sigma$ ($3.7\sigma$) at a Higgs boson
mass of 125\GeV.
Correspondingly, the signal
strength relative to the expectation for the
standard model Higgs boson is found to be $0.78 \pm 0.27$
at the Higgs boson mass of 125\GeV.
Earlier, the CDF and D0 experiments at the Tevatron reported, for $\mh
= 125\GeV$, an observed (expected) 95\% CL upper limit of 7.0 (5.7)~times the standard model
expectation~\cite{PhysRevD.88.052014}.
Using the LHC data taken at 7\TeV, the ATLAS and CMS Collaborations
reported, for $\mh=125\GeV$, observed (expected) 95\% CL upper limits of 3.7
(3.5)~\cite{Aad:2012mea} and 4.2 (3.1)~\cite{Chatrchyan:2012vp}~times the standard model
expectation, respectively.

The CMS measurements in the V\hbb and \htt
searches~\cite{Chatrchyan:2013zna,Chatrchyan:2014nva} are both consistent, within
the precision of the present data, with each other and with
the expectation for the production and decay of the standard model Higgs boson.
Hence, as a next step, a combination of these two results, requiring the
simultaneous analysis of the data selected by the two individual measurements,
is performed.
Such a simultaneous fit to both datasets accounts for all statistical and
systematic uncertainties and their correlations.
The statistical methods used for the combination were developed
by the ATLAS and CMS Collaborations in the context of the LHC Higgs Combination
Group.
The presence of the signal is assessed by performing a fit to the observed data
and using a test statistic based on the profile likelihood ratio.
Systematic uncertainties are incorporated
through nuisance parameters and are treated according to the frequentist
paradigm.
A detailed description of the methodology can be found in
Refs.~\cite{LHC-HCG-Report, Chatrchyan:2012tx} and results presented here are
obtained using asymptotic formulae~\cite{Cowan:2010st}.
To quantify an excess of events over the background expectation, the test
statistic is defined as

\begin{equation}
\label{eq:method_q0}
 q_{0} \, = \, - 2 \, \ln \frac {\mathcal{L}(\text{obs.} \, | \, b, \, \hat
 \theta_{0} ) } {\mathcal{L}(\text{obs.} \, | \, \hat \mu \cdot s + b, \, \hat
 \theta ) } ,
\end{equation}

where the likelihood, $\mathcal{L}$, appearing in the numerator is that of the
background-only hypothesis, $s$ stands for the signal expected under the standard model Higgs
boson hypothesis, $\mu$ is a signal strength modifier
introduced to accommodate deviations from standard model Higgs boson
predictions, $b$ stands for background contributions, and $\theta$ are nuisance
parameters describing the systematic uncertainties.
The value $\hat \theta_{0}$ maximises the likelihood in
the numerator under the background-only ($\mu=0$) hypothesis, while
$\hat \mu$ and $\hat \theta$ define the point at which the
likelihood reaches its global maximum in the signal-plus-background fit.
The post-fit model, obtained after the signal-plus-background fit to the data,
is used when deriving expected quantities.
The post-fit model corresponds to the parametric bootstrap described in
the statistics literature, and includes information gained in the fit regarding
the values of all parameters~\cite{Efron1979,Lee2005}.
The $p$-value for the background-only hypothesis is calculated as the tail
probability $p_{0}=P(q_{0}\geq q^\text{obs.}_{0}|b)$, where $q^\text{obs.}_{0}$
is the value of the test statistic observed in data.

While the expected Higgs boson signal in the V\hbb analysis consists exclusively
of Higgs bosons decaying to down-type fermions, that is not the case in some
event categories in the \htt analysis where a sizable contribution
from \hww decays is expected.
The contribution from \hww decays has kinematic properties
sufficiently different from those of \htt decays such that the two contributions do not
overlap.
In order to exclusively assess the fermionic decays of a
Higgs boson, the expected contribution from \hww
decays is considered as a background process.
Given the discovery of a Higgs boson
with properties as expected in the standard model and a mass close to 125\GeV,
the $\Pp\Pp\to\hww$ background contribution is taken from the
expectation for the standard model Higgs boson with a mass of 125\GeV,
including all associated uncertainties.

\begin{figure}[thb]
\centering
\includegraphics[width=\cmsFigWidth]{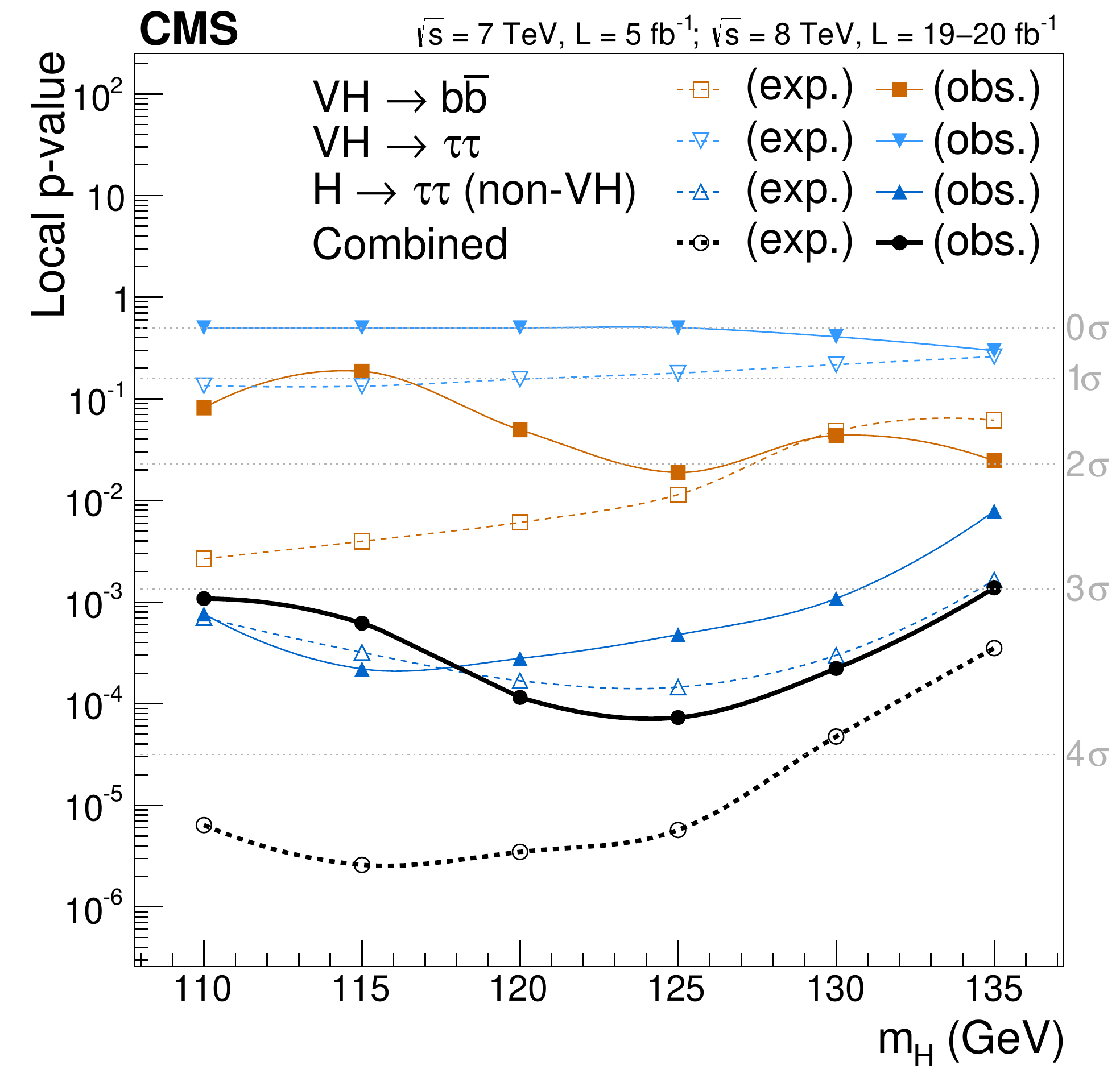}%
\caption{
The $p$-value for the background-only hypothesis as a function of the Higgs
boson mass hypothesis for both the observed data (obs.) and pseudo-data constructed using the median
expectation for the standard model Higgs boson after the signal-plus-background
fit to the data (exp.).
To help visualize the results, smooth lines connect the points at which the
probabilities are evaluated.
The \htt analysis results are broken down into a
contribution from V\htt categories, which share the production process with
the V\hbb analysis, and another contribution with all other \htt event
categories.
For every mass hypothesis tested, the non-fermionic decay
contributions are taken to be those of the standard model Higgs boson with a mass of
125\GeV.
Non-fermionic decays are considered as part of the background and consequently
do not contribute to the measurement.
}
\label{fig:significance}
\end{figure}

Figure~\ref{fig:significance} shows, as a function of the mass
hypothesis for the Higgs boson decaying to fermions, the
median expected and the observed $p$-value for the background-only hypothesis.
The expectation is calculated after having performed the
signal-plus-background fit to the observed data.
The background-only hypothesis includes the
$\Pp\Pp\to\PH\text{(125\GeV)}\to\ww$ process for every value of
\mh.
The $p$-value can be expressed in terms of Gaussian tail probabilities
and given in units of standard deviation ($\sigma$), shown in
the right vertical axis of Fig.~\ref{fig:significance}.
For all \mh values tested, the evidence against the
background-only hypothesis is found to be 3$\sigma$ or more, with a maximum of
3.8$\sigma$ for $\mh=125\GeV$, corresponding to a $p$-value of
$7.3\ten{-5}$.

\begin{figure}[thb]
\centering
\includegraphics[width=\cmsFigWidth]{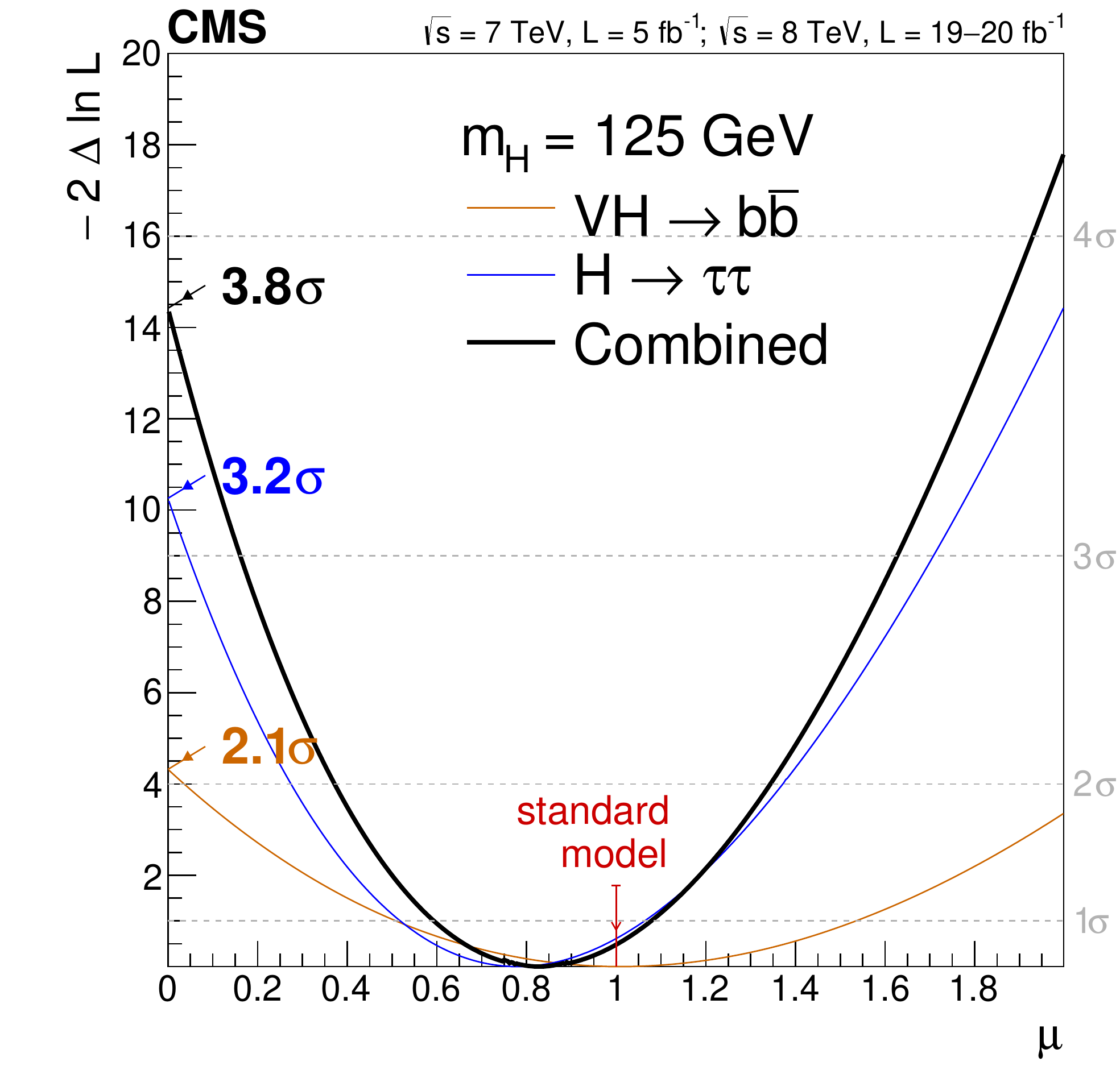}
\caption{
Scan of the profile likelihood as a function of the signal strength relative to
the expectation for the production and decay of a standard model Higgs boson, $\mu$, for
$\mh=125\GeV$.
The statistical significance against the background-only ($\mu=0$) hypothesis
is shown for the two channels and their
combination.
By definition, the expectation for the standard model Higgs boson with a mass of
125\GeV is $\mu=1$.
The non-fermionic decay contributions expected for the standard model Higgs boson
with a mass of 125\GeV are considered as part of the background and therefore
not scaled with $\mu$. }
\label{fig:nllscan}
\end{figure}

The study of the fermionic-decay contribution under the $\mh=125\GeV$ hypothesis
is particularly relevant because of the discovery of a Higgs boson with a mass
near 125\GeV in the analysis of non-fermionic decays.
Since the mass resolution in the \hbb and \htt channels is, at best, around
10\%, and the expected yields do not change significantly around $\mh=125\GeV$,
the exact choice of \mh does not affect the conclusions.
A scan of the profile likelihood is performed to estimate the best-fit signal
strength relative to the standard model expectation as well as confidence
intervals.
As described earlier, the \hww contribution is not treated as part of the signal.
The result is shown in Fig.~\ref{fig:nllscan}, from which a combined
signal strength of $0.83\pm 0.24$ with respect to the standard model expectation is
inferred for the production of a Higgs boson decaying to down-type fermions.
The statistical component of the uncertainty represents 80\%
of the total uncertainty.
This result is compatible with the expectation for the
standard model Higgs boson, as summarised in Table~\ref{tab:results}.

\begin{table}[htbp]
\centering
\topcaption{
Summary of results for the Higgs boson mass hypothesis of 125\GeV.
The $p$-values for the background-only hypothesis are
expressed in terms of one-sided Gaussian tail significances and are
provided in units of standard deviation ($\sigma$).
The expected significance is that obtained after
the fit of the signal-plus-background hypothesis to the data.
Note that the expected significance of 2.1$\sigma$ quoted
in Ref.~\cite{Chatrchyan:2013zna} for the V\hbb channel was obtained before the
fit of the signal-plus-background hypothesis to the data.
The best-fit value of the signal strength relative to
the expectation from the standard model, $\mu$,
summarises the profile likelihood scan of Fig.~\ref{fig:nllscan}.
For simplicity, uncertainties have been symmetrised.
The statistical component represents more than 80\% of the uncertainties.}
\label{tab:results}
\begin{tabular}{lccc}
\hline
Channel & \multicolumn{2}{|c|}{Significance ($\sigma$)} & Best-fit \\
\cline{2-3}
($\mh=125\GeV$) &\multicolumn{1}{|c}{ Expected} & \multicolumn{1}{|c|}{Observed} & $\mu$\\
\hline \hline

V\hbb  & 2.3 & 2.1 & $1.0\pm 0.5$ \\
\htt  & 3.7 & 3.2 & $0.78\pm 0.27$ \\
\hline
Combined & 4.4 & 3.8 & $0.83\pm0.24$ \\

\hline
\end{tabular}
\end{table}

In conclusion, the existing CMS searches for a Higgs boson decaying
into bottom quarks and $\Pgt$ leptons are consistent with the standard model
prediction of a Yukawa structure, where the fermionic couplings are
proportional to the fermion masses.
Combining the results of the CMS measurements in the \hbb and \htt decay
channels and assuming that the Higgs boson is produced as expected in the
standard model,
reveals, for the first time, strong evidence for the direct coupling of
the 125\GeV Higgs boson to down-type fermions, with an observed (expected)
significance of 3.8 (4.4) standard deviations.

\clearpage
\section*{Acknowledgements}

This letter is dedicated to our dear colleague and friend Professor Lorenzo
Fo\`{a} who passed away on January 13, 2014.
Lorenzo's contribution to CMS was unique and left an indelible impact on the
experiment in all its phases:
Lorenzo was the CMS principal referee in the CERN LHC Committee during the
approval process,
initiated the participation of many groups during the formation of the CMS
Collaboration,
provided prescient advice and guidance during his period as CERN
Director of Research,
was Chair of the CMS Collaboration Board during the critical phase of
construction,
and was the first Chair of the Thesis Award Committee enabling him to stay close
to young physicists.

We congratulate our colleagues in the CERN accelerator departments for the
excellent performance of the LHC and thank the technical and administrative
staffs at CERN and at other CMS institutes for their contributions to the
success of the CMS effort. In addition, we gratefully acknowledge the computing
centres and personnel of the Worldwide LHC Computing Grid for delivering so
effectively the computing infrastructure essential to our analyses. Finally, we
acknowledge the enduring support for the construction and operation of the LHC
and the CMS detector provided by the following funding agencies: BMWF and FWF
(Austria); FNRS and FWO (Belgium); CNPq, CAPES, FAPERJ, and FAPESP (Brazil);
MES (Bulgaria); CERN; CAS, MoST, and NSFC (China); COLCIENCIAS (Colombia); MSES
and CSF (Croatia); RPF (Cyprus); MoER, SF0690030s09 and ERDF (Estonia); Academy
of Finland, MEC, and HIP (Finland); CEA and CNRS/IN2P3 (France); BMBF, DFG, and
HGF (Germany); GSRT (Greece); OTKA and NIH (Hungary); DAE and DST (India); IPM
(Iran); SFI (Ireland); INFN (Italy); NRF and WCU (Republic of Korea); LAS
(Lithuania); MOE and UM (Malaysia); CINVESTAV, CONACYT, SEP, and UASLP-FAI
(Mexico); MBIE (New Zealand); PAEC (Pakistan); MSHE and NSC (Poland); FCT
(Portugal); JINR (Dubna); MON, RosAtom, RAS and RFBR (Russia); MESTD (Serbia);
SEIDI and CPAN (Spain); Swiss Funding Agencies (Switzerland); NSC (Taipei);
ThEPCenter, IPST, STAR and NSTDA (Thailand); TUBITAK and TAEK (Turkey); NASU
(Ukraine); STFC (United Kingdom); DOE and NSF (USA).

\bibliography{auto_generated}   

\cleardoublepage \appendix\section{The CMS Collaboration \label{app:collab}}\begin{sloppypar}\hyphenpenalty=5000\widowpenalty=500\clubpenalty=5000\textbf{Yerevan Physics Institute,  Yerevan,  Armenia}\\*[0pt]
S.~Chatrchyan, V.~Khachatryan, A.M.~Sirunyan, A.~Tumasyan
\vskip\cmsinstskip
\textbf{Institut f\"{u}r Hochenergiephysik der OeAW,  Wien,  Austria}\\*[0pt]
W.~Adam, T.~Bergauer, M.~Dragicevic, J.~Er\"{o}, C.~Fabjan\cmsAuthorMark{1}, M.~Friedl, R.~Fr\"{u}hwirth\cmsAuthorMark{1}, V.M.~Ghete, C.~Hartl, N.~H\"{o}rmann, J.~Hrubec, M.~Jeitler\cmsAuthorMark{1}, W.~Kiesenhofer, V.~Kn\"{u}nz, M.~Krammer\cmsAuthorMark{1}, I.~Kr\"{a}tschmer, D.~Liko, I.~Mikulec, D.~Rabady\cmsAuthorMark{2}, B.~Rahbaran, H.~Rohringer, R.~Sch\"{o}fbeck, J.~Strauss, A.~Taurok, W.~Treberer-Treberspurg, W.~Waltenberger, C.-E.~Wulz\cmsAuthorMark{1}
% ...
\vskip\cmsinstskip
% ...
1:~~Also at Vienna University of Technology, Vienna, Austria\\
2:~~Also at CERN, European Organization for Nuclear Research, Geneva, Switzerland\\
% ...

\end{sloppypar}
\end{document}